\documentclass[prl,twoside,twocolumn,superscriptaddress,showpacs]{revtex4}
\usepackage{graphicx}
\usepackage{amsmath}
\usepackage{amssymb}

\begin{document}

\title{Incommensurate Charge Order Phase in Fe$_2$OBO$_3$ due to Geometrical Frustration}

\author{M. Angst}
 \email[email: ]{angst@ornl.gov}
\affiliation{Materials Science and Technology Division, Oak Ridge
National Laboratory, Oak Ridge, TN 37831, USA}
\author{R.~P. Hermann}
\affiliation{Institut f\"ur Festk\"orperforschung, Forschungszentrum
J\"ulich GmbH, D-52425 J\"ulich, Germany} \affiliation{Department of
Physics, B5, Universit\'e de Li\`ege, B-4000 Sart-Tilman, Belgium}
\author{W. Schweika}
\affiliation{Institut f\"ur Festk\"orperforschung, Forschungszentrum
J\"ulich GmbH, D-52425 J\"ulich, Germany}
\author{J.-W. Kim}
\affiliation{Ames Laboratory, Ames, IA 50010, USA}
\author{P. Khalifah}
\affiliation{Department of Chemistry, University of Massachusetts,
Amherst, MA 01003, USA}
\author{H.~J. Xiang}
\author{M.-H.~Whangbo}
\affiliation{Department of Chemistry, North Carolina State
University, Raleigh, NC 27695, USA}
\author{D.-H. Kim}
\author{B.~C. Sales}
\author{D.~Mandrus}
\affiliation{Materials Science and Technology Division, Oak Ridge
National Laboratory, Oak Ridge, TN 37831, USA}

\date{\today}
\begin{abstract}
The temperature dependence of charge order in Fe$_2$OBO$_3$ was
investigated by resistivity and differential scanning calorimetry
measurements, M\"ossbauer spectroscopy, and synchrotron X-ray
scattering, revealing an intermediate phase between room temperature
and $340\,{\rm K}$, characterized by coexisting mobile and immobile
carriers, and by incommensurate superstructure modulations with
temperature-dependent propagation vector $(\frac{1}{2},0,\tau)$. The
incommensurate modulations arise from specific anti-phase boundaries
with low energy cost due to geometrical charge frustration.
\end{abstract}

\pacs{71.30.+h, 64.70.Rh, 61.10.Eq, 76.70.+y}

\maketitle

Geometrical frustration, the inability of interactions to be
simultaneously satisfied due to the geometry of the underlying
crystal lattice, in the case of magnetic interactions is known to be
responsible for incommensurate spin modulations \cite{Lawes04} and
for various exotic ground states \cite{Moessner06}. The
Coulomb-interactions leading to charge order (CO)
\cite{Wigner34,Verwey39} can also be geometrically frustrated
\cite{Seo06}. However, the effects of geometrical charge frustration
are less well understood, despite of the classical CO example
\cite{Verwey39} occurring on a frustrated lattice. Observations of
incommensurate charge modulations are typically attributed to
charge-density-wave-like Fermi surface nesting scenarios
\cite{Gruner94,ChuangMilward}, but unusually complex CO patterns
observed in spinels \cite{RadaelliHoribe}, ``devil's
staircase''-type CO modulations found in NaV$_2$O$_5$ under pressure
\cite{Ohwada01}, and the appearance of a new type of
ferroelectricity upon charge ordering in LuFe$_2$O$_4$
\cite{Ikeda05}, may be related to charge frustration.

\begin{figure}[!b]
\includegraphics[width=0.83\linewidth]{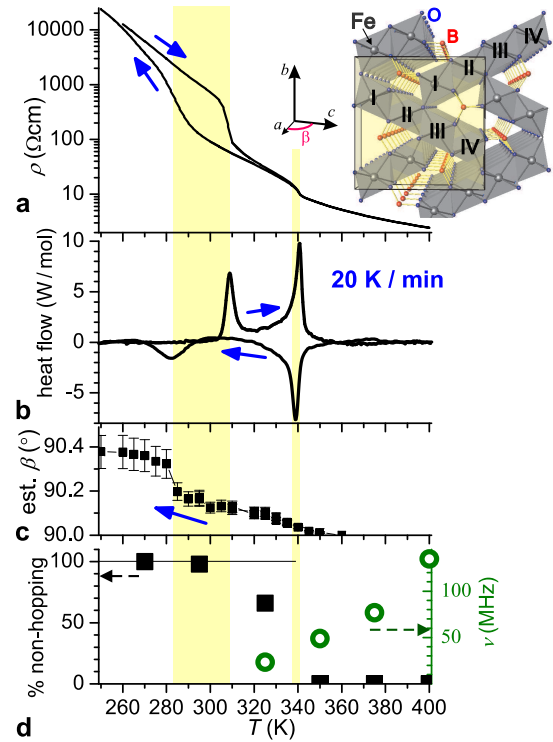}
\caption{ (Color online) Bulk physical properties of Fe$_2$OBO$_3$
indicating two phase transitions. a$-$d: $T$ dependence of
resistivity $\rho$ along $a$ (a), heat flow of differential scanning
calorimetry (b), estimated monoclinic angle $\beta$ (c), and key
parameters describing the M\"{o}ssbauer spectra (d): fraction of the
contribution to the spectra by Fe ions with no discernible electron
hopping ($\blacksquare$, left) and hopping frequency $\nu$ of the
remainder ($\bigcirc$, right). Inset in a: Crystal structure at
$355\,{\rm K}$ (after \cite{Fe2OBO3first}).} \label{Fig1}
\end{figure}

The warwickite crystal structure of Fe$_2$OBO$_3$ (Fig.\ \ref{Fig1}
inset) is prone to frustration, and we recently observed a CO
superstructure with integer valence states (clearly distinct from a
classical charge-density-wave) at low $T$ \cite{Fe2OBO3first}. Here,
we present a detailed study of the thermal evolution of the CO in
Fe$_2$OBO$_3$, establishing the presence of an unanticipated
intermediate phase \cite{note_Karen01} with coexisting mobile and
immobile carriers. Surprisingly, X-ray scattering indicates that the
superstructure is incommensurate in the intermediate phase with a
$T$ dependent propagation vector $(\frac{1}{2},0,\tau)$. We show how
the incommensuration arises from geometrical charge frustration via
the proliferation of low-energy anti-phase-boundaries, as proposed
also for {\em chemical order} in binary alloys \cite{Paxton97}.
Similar incommensurate phases may occur in other frustrated systems
with "binary order" (e.g.\ Ising spins) and thus our results not
only provide the first example of incommensurate phases with ionic
(and therefore binary) CO, but have implications for the broader
study of geometrical frustration. In particular, we propose
Fe$_2$OBO$_3$ as an ideal system to study ordering {\em dynamics} on
frustrated lattices.

We studied single crystals from the same batch as those studied in
\cite{Fe2OBO3first}. The electrical resistivity was measured with
standard four-probe geometry, using gold-sputtered contacts.
Differential scanning calorimetry (DSC) was performed, with a
Perkin-Elmer Diamond DSC, on several powdered crystals, with
temperature sweep-rates between $10$ and $50\,{\rm K}/{\rm min}$
giving consistent results. The details of M\"ossbauer spectroscopy
and electronic structure (GGA+U) calculations are as in
\cite{Fe2OBO3first}.

Synchrotron X-ray scattering was performed on two crystals with a
mosaic spread of $0.20(8)^{\circ}$ at the MU-CAT sector $6{\rm ID}B$
of the Advanced Photon Source in the Argonne National Laboratory,
with the photon energy tuned to $7.05\,{\rm keV}$. All scattered
intensities shown are normalized by an ion chamber monitor to
account for changes in the incoming beam intensity. Due to the
crystal mosaic spread and the presence of twinning, we determined
the monoclinic distortion by rotating the crystal to the nominal
($1,3,3$) position and scanning the detector angle (similar to a
powder diffraction experiment). Using the $a$, $b$, and $c$ values
obtained in \cite{Fe2OBO3first} the monoclinic angle $\beta$ was
estimated, with error bars given by the uncertainty in $a$, $b$, and
$c$. A further study was performed at $6{\rm ID}D$ with photon
energy $\sim\! 98\,{\rm keV}$ and an image plate system, using a
rocking technique as in \cite{Kreyssig07}.

Between $250$ and $400\,{\rm K}$, resistivity $\rho (T)$ and DSC
data (Fig.\ \ref{Fig1}a,b) show two separate, well defined phase
transitions on cooling(/warming) at $340$ and $280$(/$308)\,{\rm
K}$, the former of which corresponds (Fig.\ \ref{Fig1}c) to the
monoclinic-orthorhombic structural transition. The transitions
delineate low, intermediate, and high $T$ phases. Ac specific heat
had not revealed the lower transition \cite{Fe2OBO3first}, which we
attribute to its insensitivity to irreversible contributions with
hysteresis larger than the ac amplitude. In contrast, DSC is
sensitive to {\em all} contributions to the specific heat. The
estimated total entropy change through both transitions
($\sim\!0.6\,{\rm J/mol/K}/$Fe ion) still is only about 10\% of the
configurational entropy associated with perfect CO. Two likely
reasons for this will be discussed below.

\begin{figure}[t]
\includegraphics[width=0.68\linewidth]{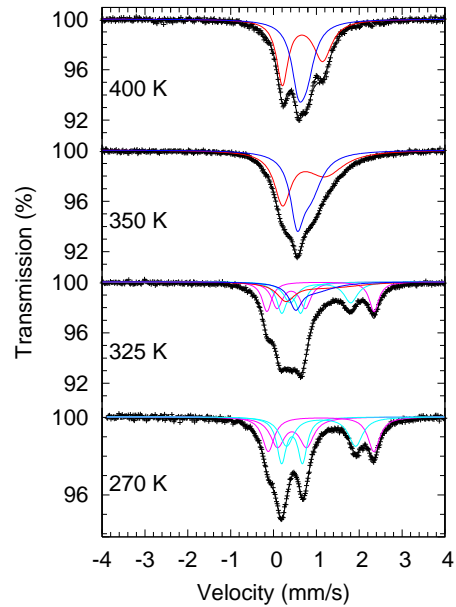}
\caption{(Color online) M{\"{o}}ssbauer spectra ($+$) at selected
$T$ on powdered Fe$_2$OBO$_3$ crystals, together with fits (black
lines). Different contributions to the fit are indicated by colored
lines. Bright blue and magenta: contributions from Fe sites with no
electron hopping (two doublets for each color correspond to two
distinct Fe valences). Blue and red: contributions from Fe sites
with non-zero electron hopping. Red and magenta doublets, with the
larger quadrupole splitting, may be assigned to the sites on the
outer chains I,IV and the blue and bright blue doublets to the sites
on the inner chains II,III \cite{AttfieldDouvalis}.} \label{FigM}
\end{figure}

M\"{o}ssbauer spectra (Fig.\ \ref{FigM}) indicate divalent and
trivalent Fe ions distributed over two structural sites, as before
\cite{AttfieldDouvalis}. However, whereas there is no discernible
hopping of electrons between Fe ions at low $T$ (hopping below
$\sim\!\! 1\,{\rm MHz}$ is not resolvable), the description of the
high $T$ spectra requires hopping with a $T$ dependent frequency
$\nu$ (Fig.\ \ref{Fig1}d, $\circ$), which remains quite low even
above $400\,{\rm K}$ \cite{note_BlumeTjon}. In the intermediate $T$
phase between the two transitions, e.g.\ at $325\,{\rm K}$,
contributions with and without electron hopping coexist (Fig.\
\ref{Fig1}d, $\blacksquare$). Isomer shifts and quadrupole
splittings are equal for both contributions, and follow a standard
$T$ dependence.

\begin{figure}[tb]
\includegraphics[width=0.99\linewidth]{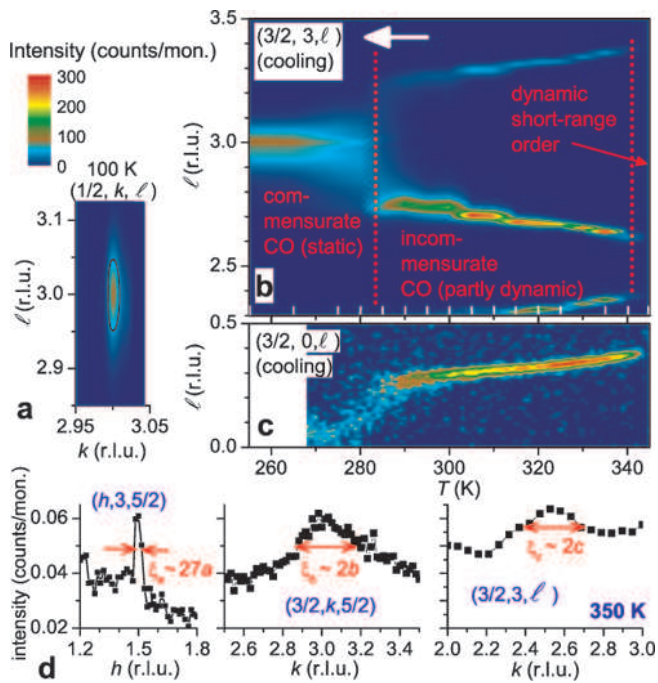}
\caption{(Color online) Commensurate and incommensurate
superstructure reflections from synchrotron X-ray scattering.
Fundamental reflections at integer $(h,k,\ell)$ typically have peak
intensities in excess of $10^4$ counts/monitor. a: Scattered
intensity at ($\frac{1}{2},k,\ell$) and $100\,{\rm K}$, showing one
of the superstructure peaks corresponding to the doubled unit cell
\cite{Fe2OBO3first}. Black line: half-maximum-intensity contour. b:
Intensity at ($\frac{3}{2},3,\ell$) vs $T$. An estimated beam
heating of $5\,{\rm K}$ was corrected. Inner white ticks indicate
$T$ at which data were taken. c: Intensity after background
subtraction at ($\frac{3}{2},0,\ell$) measured every $1\,{\rm K}$ in
a different experimental setup with an image plate system. d: $h$,
$k$, and $\ell$ scans through ($\frac{3}{2},3,\frac{5}{2}$) at
$350\,{\rm K}$. } \label{Fig3}
\end{figure}

\begin{figure}[tb]
\includegraphics[width=0.98\linewidth]{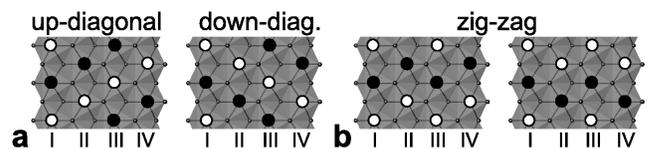}
\caption{ (Color online) Different charge order configurations
within the four-chain ribbons of the Fe$_2$OBO$_3$ structure (Fig.\
\ref{Fig1} inset). Fe$^{3+}$ and Fe$^{2+}$ are drawn as filled and
open circles, respectively. a: Diagonal charge order, the ground
state configuration. Degeneracy of the two configurations leads to
domain formation, with opposite sense of monoclinic distortion
\cite{Fe2OBO3first}. b: Alternative ``zig-zag'' CO with only
slightly higher energy than the diagonal groundstate. All other
configurations are intermediate between diagonal and zig-zag. }
\label{diagzigzag}
\end{figure}

We further characterized the three phases by X-ray scattering using
synchrotron radiation. In the low $T$ phase, weak additional
reflections (Fig.\ \ref{Fig3}a) indexing to ($h\!
+\!\frac{1}{2},k,\ell$) are observed, as expected
\cite{Fe2OBO3first}. Superstructure reflections are sharp along $h$
and $k$. Along $\ell$ they are $>\! 25\times$ broader than
fundamental reflections, and indicate a correlation length, obtained
by the inverse of the full width at half maximum, of only
$\xi_c\!\sim\! 12\, c$. This suggests differently ordered domains
(likely up and down diagonal CO \cite{Fe2OBO3first}, see Fig.\
\ref{diagzigzag}a) of size large along $a$ and $b$, but small along
$c$. These micro-domains correspond to an imperfect overall CO,
which is one reason for the lower-than-expected entropy. In the
intermediate phase superstructure reflections are sharp along $\ell$
and are split, corresponding to an incommensurate modulation, with
propagation vector ($\frac{1}{2},0,\tau$), and $\tau (T)$ varying
from about $0.4$ at $340\,{\rm K}$ to $0.2$ at $280\, {\rm K}$
(Fig.\ \ref{Fig3}b). Above $340\,{\rm K}$ a long-range ordered
superstructure no longer exists, but very weak and broad reflections
with ($h\! +\!\frac{1}{2},k,\ell\!+\!\frac{1}{2}$) index (Fig.\
\ref{Fig3}d) indicate persistent short-range correlated
fluctuations, with correlations mainly along $a$, the
chain-direction. The correlations are not static, but the dynamics
is relatively slow (Fig.\ \ref{Fig1}d). This dynamic
short-range-order likely also contributes to the ``missing
entropy''. To test the incommensurability of the modulations in the
intermediate temperature phase, data were collected in $1\,{\rm K}$
intervals  with high-energy X-rays. No indications for any
``lock-in'' to commensurate values are visible in Fig.\ \ref{Fig3}c,
thus the modulations are truly incommensurate rather than forming a
``devil's staircase'' as in \cite{Ohwada01}.

\begin{figure}[tb]
\includegraphics[width=0.91\linewidth]{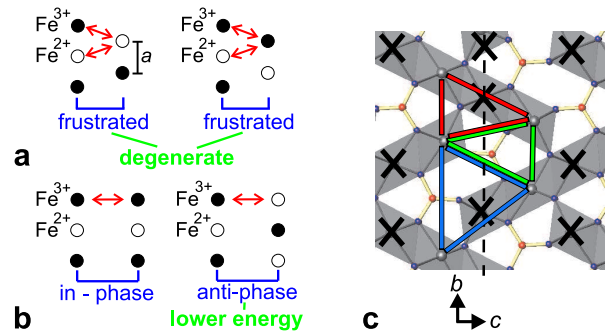}
\caption{ (Color online) Impact of geometrical frustration on the
charge order. a,b: Phase-relationship between two charge-ordered
chains: For chains offset by $\frac{a}{2}$ (a), inter-chain
electrostatic interactions are frustrated, and two configurations
are degenerate in energy. Without offset (b), the anti-phase
configuration has lower energy. c: $bc$ projection of the
Fe$_2$OBO$_3$ structure. The arrangement of chains leads to
geometrical frustration even for interactions (thick lines) between
chains within one sublattice, with possible phase relationships as
in b (the other chains are crossed out). Dashed line: possible
anti-phase-boundary (see text).} \label{frust}
\end{figure}

To elucidate the micro-domain formation at low $T$, and the
incommensurate CO at intermediate $T$, we consider the energy of
various CO configurations. The electrostatic energy is minimized by
having as few same-valence nearest neighbors as possible. This
requires strictly alternating valences along $a$ within each chain
and establishes ``ordered chains'' as the basic CO unit, consistent
with the correlations in the fluctuation regime (Fig.\ \ref{Fig3}d).
Geometrically, the 2D lattice of chains consists of two sublattices
offset from each other by $\frac{a}{2}$ ($1.6\,{\rm {\AA}}$): One
sublattice comprises chains numbered I and III in Fig.
\ref{diagzigzag} and Fig.\ \ref{Fig1} inset, the other chains II and
IV. The electrostatic interactions between any two chains not
belonging to the same sublattice are geometrically frustrated (Fig.\
\ref{frust}a). Therefore, flipping all valences in all chains
(changing their phase) of one sublattice does not change the energy.
Thus, the two configurations of Fig.\ \ref{diagzigzag}a are
degenerate in energy, as are the two of Fig.\ \ref{diagzigzag}b.
This twofold degeneracy is lifted by a monoclinic distortion (ribbon
tilting), and hence the system will distort, thereby gaining energy
\cite{Jahn37} However, according to our GGA+U calculations the
energy gain associated with this distortion is very small, $\sim\!
0.2\%$ of the overall CO gain for diagonal CO (Fig.\
\ref{diagzigzag}a), insufficient to prevent the facile formation of
domains.

Neglecting the small monoclinic distortion there is no net
interaction between the two sublattices so that further
considerations can be restricted to one of them. Any two chains of
the sublattice are either in-phase or anti-phase (Fig.\
\ref{frust}b) with a lower electrostatic energy if they are
anti-phase. Nevertheless, this does not remove all frustration
because there is additional geometrical frustration even within one
sublattice, due to the arrangement of chains projected in the $bc$
plane (Fig.\ \ref{frust}c): several ``frustration triangles'' are
readily visible (highlighted in red, green, blue). Although this
frustration is not exact (triangle sides have unequal length),
various configurations are expected to be very close in energy.
Indeed, while our GGA+U calculations established a diagonal CO
(Fig.\ \ref{diagzigzag}a) ground state \cite{Fe2OBO3first}, they
also show that, e.g., the total energy gain due to zigzag CO (Fig.\
\ref{diagzigzag}b) is only $1.9$\% \cite{note_Xiang} smaller than
the one due to diagonal CO, despite having more chains in-phase
rather than anti-phase.

Due to this near-degeneracy of  configurations with different phase
relationships, resulting from a hierarchy of geometrical
frustration, the energy cost of creating an anti-phase boundary in
the $ab$ plane (a possible location is indicated in Fig.\
\ref{frust}c) between up and down diagonal CO domains (Fig.\
\ref{diagzigzag}a) is small, explaining the small $\xi_c$. In
contrast, there is no geometrical frustration within a chain,
leading to a much larger $\xi_a$. $\xi_b$ is large also due to the
association of different CO domains with monoclinic distortions of
opposite sense. A boundary in the $ac$ plane between domains of
different monoclinic angle $\beta$ is not possible (short of
cleaving the crystal), because the lattices will not match. In
contrast, there is no problem matching domains with different
$\beta$ in $ab$ or $bc$ planes.

Incommensurate CO in classical charge-density-wave systems
\cite{Gruner94} and colossal magnetoresistance manganites
\cite{ChuangMilward} has been associated with Fermi-surface nesting.
However, because Fe$_2$OBO$_3$ is far from metallic (Fig.\
\ref{Fig1}a,d), a nesting scenario can not account for the
incommensurate modulations in the intermediate phase. A better
analogy to our case of essentially perfect ionic CO is found in the
{\em chemical} order of ions in binary alloys. In fcc-structured
alloys, such as Cu-Au, incommensurate phases between chemically
ordered (commensurate) and disordered phases are often observed, and
can be explained as a modulation with statistical period arising
from anti-phase boundaries easily created due to degeneracy and
frustration of the bonds on triangular networks \cite{Paxton97}.
Given that in the intermediate phase of Fe$_2$OBO$_3$ the
incommensurate modulation is in the $c$ direction and the anti-phase
boundaries $\perp \! c$ are the ones with a small energy cost, we
deduce that a similar scenario applies.

The strong and continuous $T$ dependence of the propagation vector
then implies that the boundaries move easily in the intermediate
phase, and thermal excitations should cause them to fluctuate.
Fluctuating boundaries reduce the monoclinic distortion (as
observed, Fig.\ \ref{Fig1}c) and they imply that a fraction of Fe
ions, those near the boundaries, change their valences. Fe ions
change their valence by electron hopping, providing a natural
explanation for the presence, in the intermediate $T$ phase, of both
contributions with and without electron hopping in the M\"{o}ssbauer
spectra (Fig.\ \ref{Fig1}d). The coexistence of mobile and immobile
electrons, reminiscent of nano-scale inhomogeneities prominent in
manganites and cuprates \cite{MathurDagotto}, but coherently ordered
in Fe$_2$OBO$_3$, further stabilizes the intermediate phase as a
compromise between CO and delocalizing tendencies influenced by the
geometrical frustration (without which the transition to a charge
ordered state likely would occur at much higher $T$
\cite{notefrust}). Similar compromise phases may occur in other
correlated electron systems with no chemical disorder.

In summary, we discovered an intermediate $T$ phase of CO in
Fe$_2$OBO$_3$, which is characterized by the coexistence of mobile
and immobile carriers, and by an incommensurate superstructure, the
latter of which can be explained by specific anti-phase boundaries
created easily due to geometrical charge frustration, similar to
incommensurate {\em chemical} order. It should be interesting to
further scrutinize the dynamics \cite{Seo06} of this model-system of
incommensuration from geometrical charge frustration.

We thank A. Payzant, D.~S. Robinson, J. Tao, S. Nagler, F.
Grandjean, J.~W. Brill, K.~Conder, and R.~Puzniak for assistance and
discussions. Research at ORNL sponsored by the Division of Materials
Sciences and Engineering, Office of Basic Energy Sciences (OS), US
Department of Energy (DOE), under contract DE-AC05-00OR22725 with
ORNL, at NCSU and synchrotron work at the Advanced Photon Source
(6IDB/D) by OS, DOE (contracts DE-FG02-86ER45259 and
W-31-109-Eng-38). Work at ULg: FNRS credit 1.5.064.05.

\newcommand{\noopsort}[1]{} \newcommand{\printfirst}[2]{#1}
  \newcommand{\singleletter}[1]{#1} \newcommand{\switchargs}[2]{#2#1}

\end{document}